\documentclass[article]{IEEEtran} 
\usepackage{xurl} 
\usepackage{hyperref}
\usepackage{amsmath,amsfonts}
\usepackage{array}
\usepackage[noadjust]{cite}
\usepackage{textcomp}
\usepackage{stfloats}
\usepackage{url}
\usepackage{verbatim}
\usepackage{graphicx}
\usepackage{balance}
\usepackage{tikz}
\usetikzlibrary{fadings}
\usetikzlibrary{patterns}
\usetikzlibrary{shadows.blur}
\usetikzlibrary{shapes}   
\usepackage{siunitx}
\DeclareSIUnit\mt{\milli\tesla} 
\usepackage[utf8]{inputenc}
\usepackage[english]{babel}
\usepackage{xurl} 
\usepackage[linesnumbered,ruled,lined]{algorithm2e}
\usepackage{array}
\usepackage[noadjust]{cite}
\usepackage{textcomp}
\usepackage{url}
\usepackage{verbatim}
\usepackage{graphicx}
\usepackage{balance}
\usepackage{multirow} 
\usepackage{setspace}
\usepackage{graphicx}
\usepackage{url}
\usepackage{subcaption}
\usepackage{amsmath}
\usepackage{algcompatible}
\usepackage{array}
\usepackage{amsthm}
\usepackage{adjustbox}

\usepackage{txfonts}  
\usepackage{stfloats} 

\usepackage{color}

\usepackage{pgfplots}
\usepackage{tikz}
\usetikzlibrary{fadings}
\usetikzlibrary{patterns}
\usetikzlibrary{shadows.blur}
\usetikzlibrary{shapes} 
\usepackage{bbm}
\let\oldnl\nl
\newcommand{\nonl}{\renewcommand{\nl}{\let\nl\oldnl}}
\newcolumntype{L}[1]{>{\raggedright\let\newline\\\arraybackslash\hspace{0pt}}m{#1}}
\newcolumntype{C}[1]{>{\centering\let\newline\\\arraybackslash\hspace{0pt}}m{#1}}
\newcolumntype{R}[1]{>{\raggedleft\let\newline\\\arraybackslash\hspace{0pt}}m{#1}}
\algnewcommand\KwEvl{\textbf{Evaluation:}}
\usepackage{booktabs}
\usepackage{makecell}
\usepackage{amsmath}
\usepackage{siunitx}
\usepackage{caption} 

\pgfplotsset{compat=1.18}

\begin{document}

\title{Explainable and Robust Millimeter Wave \\ Beam Alignment for AI-Native 6G Networks
\author{Nasir Khan, Asmaa Abdallah, Abdulkadir Celik, Ahmed M. Eltawil, and Sinem Coleri}

\thanks{N. Khan and S. Coleri are with the Department of Electrical and Electronics Engineering, Koç University, Istanbul, Turkey. A. Abdallah, A. Celik, and A. M. Eltawil are with the Computer, Electrical, and Mathematical Sciences and Engineering (CEMSE) Division, King Abdullah University of Science and Technology (KAUST), Thuwal 23955, Saudi Arabia. N. Khan and S. Coleri acknowledge support from the Scientific and Technological Research Council of Turkey under Grant \#119C058 and Ford Otosan.}}


\maketitle

\begin{abstract}
Integrated artificial intelligence (AI) and communication has been recognized as a key pillar of 6G and beyond networks. In line with AI-native 6G vision, explainability and robustness in AI-driven systems are critical for establishing trust and ensuring reliable performance in diverse and evolving environments. This paper addresses these challenges by developing a robust and explainable deep learning (DL)-based beam alignment engine (BAE) for millimeter-wave (mmWave) multiple-input multiple-output (MIMO) systems. The proposed convolutional neural network (CNN)-based BAE utilizes received signal strength indicator (RSSI) measurements over a set of wide beams to accurately predict the 
best narrow beam for each UE, significantly reducing the overhead associated with exhaustive codebook-based narrow beam sweeping for initial access (IA) and data transmission. To ensure transparency and resilience, the Deep k-Nearest Neighbors (DkNN) algorithm is employed to assess the internal representations of the network via nearest neighbor approach, providing human-interpretable explanations and confidence metrics for detecting out-of-distribution inputs. Experimental results demonstrate that the proposed DL-based BAE  exhibits robustness to measurement noise, reduces beam training overhead by 75\% compared to the exhaustive search while maintaining near-optimal performance in terms of spectral efficiency. Moreover, the proposed framework improves outlier detection robustness by up to 5$\times$ and offers clearer insights into beam prediction decisions compared to traditional softmax-based classifiers.
\end{abstract}


\section{Introduction}
\IEEEPARstart{T}{he} International Telecommunication Union’s IMT-2030 framework identifies ``Integrated artificial intelligence (AI) and Communication'' as one of the core directions for 6G networks. This shift toward AI-driven solution necessitates greater transparency in decision-making processes, where explainability is pivotal in establishing trust in AI systems, allowing network operators and engineers to understand, validate, and troubleshoot the decisions made by deep learning (DL) models. Additionally, robustness against out-of-distribution inputs and adversarial attacks is crucial to ensure reliable performance in diverse and evolving environments. These two factors—explainability and robustness—are essential to fulfilling the broader goals of AI-native 6G networks, where AI is not just an enhancement but a foundational component integrated into communication systems \cite{our_work}.

Among their various wireless communication use cases, DL models especially offer promising solutions to address the challenges posed by massive multiple-input multiple-output (mMIMO) millimeter wave (mmWave) systems, which rely on efficient beam alignment to determine the optimal beam pair between base stations (BS) and user equipment (UE) \cite{surveyBA1}. 
In the current 5G standard,  beam alignment strategies are employed wherein the BS sweeps beams using reference signals, while the UE measures the received signal strength indicators (RSSIs) and reports the strongest beam back to the BS \cite{tutorialBA}. Standard approaches often use quantized beams, which distribute energy across the angular space via codebooks, such as discrete Fourier transform (DFT) codebooks. While DFT codebooks ensure broad coverage, they often lack the granularity needed for precise beam alignment. To address this, oversampled DFT (O-DFT) codebooks provide finer granularity at the cost of increased beam training overhead, as more beams need to be evaluated during the alignment process. In general, beam sweeping at the BS/UE  is cyclically executed via an exhaustive search to refresh and maintain continuous beam alignment, where the feedback communication overhead emerges as a critical bottleneck.

To address the challenges with non-AI beam alignment methods, DL-based solutions have attracted great interest, enabling learning from data and adapting to dynamic conditions \cite{BAE5G-Advanced}.
By incorporating real-world data, such as radar measurements \cite{CI1}, camera images \cite{camera}, and global navigation satellite system (GNSS) coordinates \cite{CI0}, DL-based systems can predict optimal beams with greater speed and accuracy. However, while DL-aided approaches have demonstrated significant improvements in beam alignment efficiency \cite{ CI1, camera, CI0}, a key challenge remains: the lack of interpretability in the decision-making process and the vulnerability to out-of-distribution inputs.
Radio engineers are often interested in mapping
data inputs, algorithm design, to the projected wireless
key performance indicators, where promoting explainability and robustness of DL-based solutions becomes necessary for automatic decision-making systems \cite{hamon2020robustness}. 
Ensuring that DL-based solutions are resilient to outliers and transparent in their decision-making is essential for standardization and commercial deployment.

This paper proposes a robust and explainable DL-based beam alignment engine (BAE) to predict mmWave beams during the initial access (IA) process with minimal beam sweeping overhead. The proposed solution is a convolutional neural network (CNN)-based beam predictor that utilizes RSSI feedback of a finite set of sensing beams (i.e., DFT codebook) to accurately predict the optimal narrow beams from the O-DFT codebook for IA and data transmission. The developed solution significantly enhances beam training efficiency by eliminating the need for exhaustive searches within the O-DFT codebook. To further add confidence, interpretability, and robustness in beam predictions, we employ the Deep k-Nearest Neighbors (DkNN) algorithm \cite{Deepknn}, which evaluates the internal representations of the neural network during test time. This provides a reliable measure of prediction credibility by examining how well test inputs align with the model’s training data. Our model-agnostic framework not only enhances interpretability but also effectively identifies out-of-distribution or outlier inputs, improving the system’s resilience to adversarial attacks and ensuring robust, reliable performance. Numerical results demonstrate that the proposed DL-aided BAE reduces the beam training overhead by 75\% compared to an exhaustive search in the O-DFT codebook while maintaining near-optimal accuracy. Additionally, by leveraging the DL model’s structure and conformity checks, the framework improves outlier detection robustness by up to 5$\times$ and provides clearer insights into beam prediction decisions compared to traditional softmax-based classifiers.

The rest of the paper is organized as follows. Section \ref{sec:system} describes the system model and the beam alignment problem formulation. Section \ref{sec:DL} presents the details of the DL-based solution for beam alignment and the proposed DkNN-based explainable
and robust beam selection framework. Section \ref{sec:simulation}  evaluates the performance of the proposed solution strategy. Finally, conclusions and future research directions are provided in Section \ref{sec:conclusion}.

\section{System Model and Problem Formulation} \label{sec:system}
\subsection{System Model}
We consider a downlink mmWave communication system, where the BS features a uniform linear array (ULA) with $N_{\mathrm{BS}}$ antenna elements to communicate with $N_\mathrm{U}$ single-antenna
UEs.  We concentrate on the scenario of multi-user beamforming, where the BS communicates with very UE using only a single stream. The channel from the BS to the  $\mathrm{UE}_u$ can be expressed based on geometric channel modeling as

\begin{equation}
    \mathbf{h}_u = \sum_{l=1}^{L} \alpha_{u,l} \mathbf{b}\left(\phi_{u,l}\right),
\end{equation}
where  $L$  denotes the number of paths, $\alpha_{u,l}$ represents the complex path gain for the $l$-th path, $\phi_{u,l} $ is the angle of departure for the $l$-th path, and $\mathbf{b}\left(\phi_{u,l}^{}\right)$  represents the array response vector, which is given by 
\begin{equation}
    \mathbf{b}(\phi_{u,l}) = \frac{1}{\sqrt{N_{\mathrm{BS}}}} \left[1, e^{j \frac{2 \pi}{\lambda} d \sin(\phi_{u,l})}, \dots, e^{j (N_{\mathrm{BS}} - 1) \frac{2 \pi}{\lambda} d \sin(\phi_{u,l})}\right]^T,
\end{equation}
where $\lambda$ is the signal wavelength, and $d=\frac{\lambda}{2}$ denotes the antenna spacing. To mitigate the hardware cost and power consumption of a fully digital system, we adopt analog-only beamforming where the BS has a single common transmit/receive radio frequency (RF) chain shared by $N_{\mathrm{BS}}$ antennas. Hence, the beamforming vector designed for the BS is given by
\begin{equation}
\mathbf{w}=\frac{1}{\sqrt{N_{\mathrm{BS}}}}\left[e^{j \varphi_1}, \ldots, e^{j \varphi_i}, \ldots, e^{j \varphi_{N_{\mathrm{BS}}}}\right]^{\top} \in \mathbb{C}^{N_{\mathrm{BS}} \times 1},
\end{equation}
where $\varphi_i$ is the phase shift of $i$-th antenna element.
We assume that the BS adopt a beamforming codebook $\mathbf{W}=$ $\left\{\mathrm{\mathbf{w}}_1, \ldots, \mathrm{\mathbf{w}}_Q\right\} $ incorporating $Q$ pre-defined beamforming vectors.  The vectors in $\mathbf{W}$ satisfy $\left\|\mathrm{\mathbf{w}}_q\right\|^2=1, \forall q \in\{1, \ldots, Q\}$  to accommodate the constant-modulus constraint of analog beamforming architecture. 

During the beam sweeping process, the BS periodically
transmits symbols  $s_w \in \mathbb{C}$ to the UEs through the beams defined by the matrix $\mathbf{W} \in \mathbb{C}^{N_{\mathrm{BS}} \times Q} $.
Following the beam sweeping process, the complex received signal at the $\mathrm{UE}_u$, using the $q$-th beamforming vector, can be expressed as

\begin{equation}
   {r}_{\mathrm{RSSI}, u} = \sqrt{P_{\mathrm{BS}}} \mathbf{h}_{u}^H \mathbf{w}_{q} s_{w} + {z}_{\mathrm{RSSI}, u},
\end{equation}
where $P_{\mathrm{BS}}$ denotes the BS transmit power, $\mathbf{h}_u \in \mathbb{C}^{N_{\mathrm{BS}} \times 1}$ is the channel vector, and ${z}_{\mathrm{RSSI}, u}$ represents the additive complex noise with power $\sigma_z^2$. Then, with unit-power transmitted symbols, the signal-to-noise ratio (SNR) at  the $\mathrm{UE}_u$ can be expressed as
\begin{equation}
    \text{SNR}_u = \frac{P_{\mathrm{BS}} \left| \mathbf{h}_{u}^H \mathbf{w}_q \right|^2}{\sigma_z^2}.
\end{equation}

For a given BS-UE pair, the optimal beam index $ q_{u}^*$ can be identified by selecting the beam that maximizes the SNR:
\begin{equation} \label{beam_selection}
    q_{u}^* = \underset{q_u \in \left\{1,2,\dots,Q\right\}}{\arg \max} \left( \frac{\left| \mathbf{h}_{u}^H \mathbf{w}_q \right|^2 P_{\mathrm{BS}}}{\sigma_z^2} \right) = \underset{q_u \in \left\{1,2,\dots,Q\right\}}{\arg \max} \left( \left| \mathbf{h}_{u}^H \mathbf{w}_q \right|^2 \right).
\end{equation}
Identifying $q_{u\mathbf{}}^*$ through exhaustive beam search over $Q$ codewords results in significant beam training overhead, which will be addressed by the proposed solution next.

\section{Deep learning framework for beam alignment} \label{sec:DL} 
This section describes the details of the DL-based beam alignment method and the proposed DkNN-based explainable and robust beam selection
framework.
\subsection{DL-based Beam Alignment}
To address the beam training overhead
caused by exhaustive search and resulting feedback communication overhead, the proposed solution leverages the RSSI values over a set of compact wide sensing beams $M_{\mathrm{w}}$ (i.e., beams from DFT codebook), to select the best narrow beam from the O-DFT codebook, thus, avoiding exhaustive search over the O-DFT codebook. For beam-sweeping, the BS transmits the pilot signals over a smaller set of beams, each at a separate time slot.  All UEs connected to the BS measure and report the RSSI values of the $M_{\mathrm{w}}$ beams. It is assumed that beam sweeping, measurement, and reporting occur within the coherence time during which the channel remains constant. The reported beam sweeping results for  UE  $u$ can be written as
\begin{equation}\label{rp}
\mathbf{x}_u=\left[\left|\left[r_{\mathrm{RSSI}, u}\right]_1\right|^2 \cdots\left|\left[r_{\mathrm{RSSI}, u}\right]_{M_{\mathrm{w}}}\right|^2\right]^T ,
\end{equation}
where $\left[r_{\mathrm{RSSI}, u}\right]_i=\sqrt{P_{\mathrm{BS}}} \mathbf{h}_u^H \mathbf{w}_is_w+z_i$ is the received signal using the $i$-th sensing beam, $\forall i \in M_{\mathrm{w}}$.

In the DL-based approach, this beam selection problem is formulated as a classification task. Particularly, the reported beam sweeping results from (\ref{rp}) are fed as input to the deep neural network (DNN)-based beam classifier denoted by $f\left(. \hspace{0.05cm}; {\boldsymbol{\theta}}\right): \mathcal{X} \rightarrow \mathbb{R}^Q$, where $\mathcal{X} \in  \mathbb{R}^{M_{w}} $ is the set of DNN's input corresponding to the RSSI values over $M_{\mathrm{w}}$ beams  and $\boldsymbol{\theta}$ denotes the DNN's weight parameters vector. The beam classifier outputs the  posterior probability distribution $\widehat{\mathbf{q}}=  f_q(\mathbf{x},\boldsymbol{\theta}) $ of each narrow beam in $Q$
being the optimal beam. 
The true beam index is represented as a one-hot vector $ \mathbf{q}^\star \in \mathbb{R}^Q $, and the loss function is formulated as the cross-entropy  between the predicted and true distributions: 
\begin{equation}\label{loss_func}
\mathcal{L}(\mathbf{x}_u,\mathbf{q}^\star,\boldsymbol{\theta} )=   - \sum_{i=1}^{Q} q_i^\star \log \widehat{q}_i,
\end{equation}
where $i$ is the index representing each possible beam, $q_i^\star$ and $\widehat{q}_i$ denote the true one-hot encoded vector and the predicted probability for beam $i \in Q$, respectively.

While the aforementioned  DL-aided approach reduces beam alignment overhead, it still faces issues with robustness and reliability in confidence estimates. DNN models often struggle to quantify prediction confidence due to their black-box nature, making it difficult to trust their decisions in real-world scenarios, where inaccurate beam selection can significantly degrade performance. Next, we propose a framework analyzing the internal representations of the neural networks to strengthen the explainability and robustness of its beam predictions.

\subsection{Proposed Explainable and Robust Beam Classifier Framework}

To quantify the confidence in the beam prediction task,  we utilize the DkNN algorithm from \cite{Deepknn}, which incorporates concepts
from conformal prediction to analyze the internal representations generated by the DNN/CNN during testing. This method identifies inconsistencies with training observations, thus providing a more reliable measure of confidence than the softmax-based prediction typically used to estimate the model’s confidence. The DkNN algorithm is model-agnostic and can be applied to any pre-trained DL model with manageable computational complexity.

\begin{figure}{}
    \centering
    \includegraphics[width=0.8\linewidth]{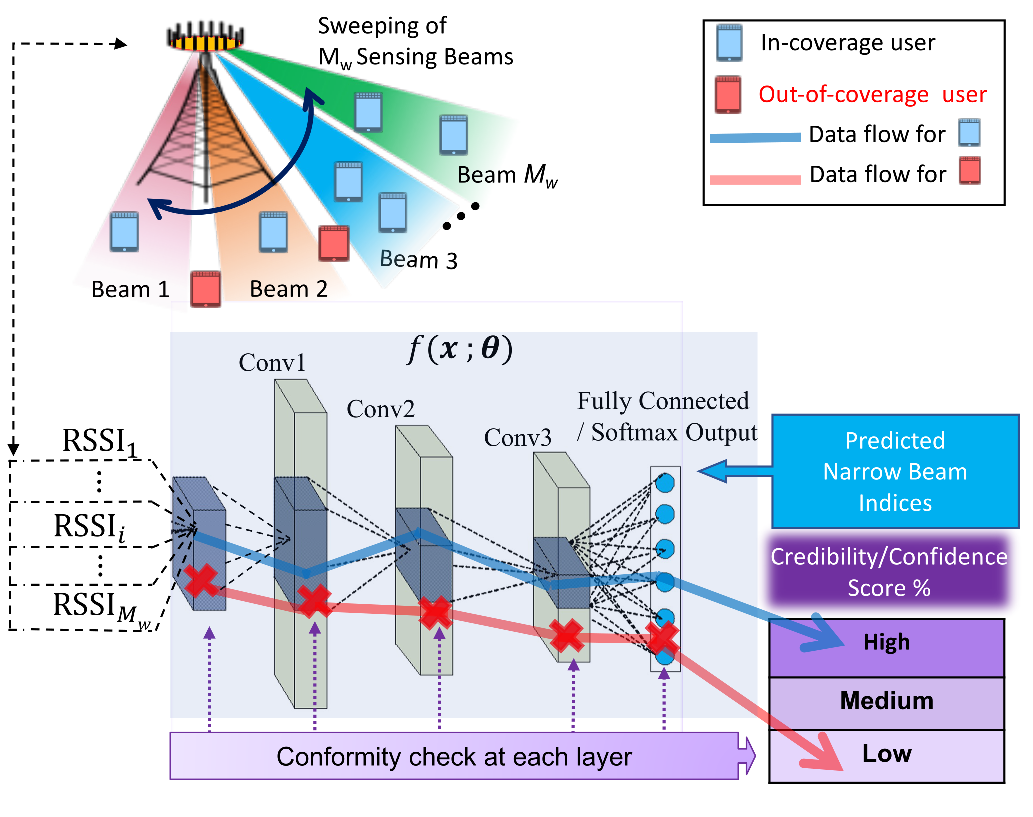}
    \caption{DkNN-based credibility assessment of the proposed beam alignment framework. }
    \label{CNN_flow}
\end{figure}

Let $f(\mathbf{x}_u; \boldsymbol{\theta})$ be a pre-trained DNN-based beam classifier composed of  $L$ layers, where each layer is indexed by $\eta$, with  $\eta \in \{0, \dots, l-1\}$ as shown in Figure \ref{CNN_flow}.  After training the DNN for beam selection, the algorithm records the output of each layer $ f_\eta $ for every training point, thus obtaining a per-layer representation of the training data paired with their respective labels. This per-layer representation is used to build a nearest neighbor classifier in the space defined by each layer $l$ to create a representation of the training data at each layer. 
To efficiently identify nearest neighbors in the high-dimensional spaces produced by these layers, we use locality-sensitive hashing (LSH) \cite{andoni2015practical}, which finds similar representations based on cosine similarity. For a given test input $\hat{\mathbf{x}}_u$, we pass it through the DNN to obtain the layer outputs $ f_\eta(\hat{\mathbf{x}}_u) $, then apply LSH to find the $ k $ nearest neighbors from the training data for each layer's representation. The labels associated with these nearest neighbors are collected for each layer, and these multi-sets of labels are used to compute the final beam selection prediction through a non-conformity check.

For a given test input $\hat{\mathbf{x}}_u$, the non-conformity of a prediction is defined as the number of nearest neighboring representations found  in training data whose label is different from the candidate label $j \in \left\{1,2,\dots,Q\right\}$, mathematically expressed as:
\begin{equation}\label{non_conformity}
\varrho(\hat{\mathbf{x}}_u, j)=\sum_{\eta \in 1, \dots, l }\left|\left\{i \in \Omega_\eta: i \neq j\right\}\right|,
\end{equation}
where $\Omega_\eta$ is the multi-set of labels for the training points whose representations are closest to the test input's at layer $\eta$, and the operator $\left|\cdot\right|$ denotes the cardinality of a set, which is the number of elements contained within it.

Before making beam predictions, we compute the nonconformity of a labeled calibration dataset $ ({X}^c, {Q}^c)$, sampled from the same distribution as the training data $ ({X}, {Q})$ but not used for training. The nonconformity values are defined as $\mathcal{C}$ =$ \left\{\varrho(\hat{\mathbf{x}}_u, {q}) : (\hat{\mathbf{x}}_u, q) \in ({X}^c, {Q}^c)\right\}$, and are then compared to the test input’s nonconformity score $\varrho(\hat{\mathbf{x}}_u, j)$ for each candidate beam label $ j $. For input $\hat{\mathbf{x}}_u$ with label $ j $, we calculate the empirical $ p $-value, which represents the fraction of calibration nonconformity scores larger than the test input’s score,  as: 
\begin{equation}
p_j(\hat{\mathbf{x}}_u) = \frac{|\{\varrho \in \mathcal{C} : \varrho \geq \varrho(\hat{\mathbf{x}}_u, j)\}|}{|\mathcal{C}|}.
\end{equation}

The predicted beam label is the one with the highest $p$-value, as given by \begin{equation}
    \text{Prediction} = \arg \max_{j \in \left\{1,2,\dots,Q\right\}} p_j(\hat{\mathbf{x}}_u).
    \label{eq:prediction}
\end{equation}
Confidence is defined as one minus the second-highest $p$-value, i.e., the probability that any label other than the prediction is true, as given by
\begin{equation}
    \text{Confidence} = 1 - \max_{j \in \left\{1,2,\dots,Q\right\}, j \neq \text{prediction}} p_j(\hat{\mathbf{x}}_u).
    \label{eq:confidence}
\end{equation}
The prediction credibility is the $p$-value of the predicted beam label, which measures the degree to which the test input conforms to the training data (equation \ref{eq:credibility}). 
\begin{equation}
    \text{Credibility} = \max_{j \in\left\{1,2,\dots,Q\right\}} p_j(\hat{\mathbf{x}}_u).
    \label{eq:credibility}
\end{equation}

\begin{algorithm}

\caption{DkNN-based credible beam selection}
\label{DkNN_algo}\small
\KwIn{training data $({X}, {Q})$, calibration data $({X}^c, {Q}^c)$}
\KwIn{trained neural network $f$ with $L$ layers}
\KwIn{number $\Tilde{k}$ of neighbors}
\KwIn{test input $\hat{\mathbf{x}}_u$}
\For{$\eta = 1$ to $L$}{
    $\Pi \leftarrow \Tilde{k}$ points in ${X}$ closest to $\hat{\mathbf{x}}$ found using LSH tables\;
    $\Omega_\eta \leftarrow \left\{{q}_i : i \in \Pi\right\}$ 
    }
$\mathcal{C} = \left\{\varrho(\mathbf{x}_u, {q}) : (\mathbf{x}_u, {q}) \in \left({X}^c, {Q}^c\right)\right\}$ \;
\For{$j = 1$ to $Q$}{
    $\varrho(\hat{\mathbf{x}}_u, j) \leftarrow \sum_{\eta = 1}^{L} \left|\left\{ i \in \Omega_\eta : i \neq j \right\}\right|$\;
    \BlankLine
    $p_j(\hat{\mathbf{x}}_u) \leftarrow \frac{|\{\varrho \in \mathcal{C} : \hspace{0.1cm}\varrho \geq \varrho(\hat{\mathbf{x}}_u, j)\}|}{|\mathcal{C}|}$
}
Calculate prediction, confidence, and credibility using equations \eqref{eq:prediction}, \eqref{eq:confidence}, and \eqref{eq:credibility}\;
\Return prediction, confidence, and credibility\;
\end{algorithm}

The proposed DkNN-based beam classification algorithm is summarized in Algorithm \ref{DkNN_algo}.
To evaluate the model’s resilience against
adversarial/outlier inputs using Algorithm \ref{DkNN_algo}, we generate the out-of-training
data (adversarial) examples by the Fast
Gradient Sign Method (FGSM) \cite{adv_FGSM} that aims
to manipulate the inputs to a beam classifier by perturbing them in the direction that maximizes the loss function in \ref{loss_func} with respect to
the true beam labels. To generate the adversarial example, the FGSM  computes the perturbations as
$
\boldsymbol{\delta}_u = \epsilon \operatorname{sign}\left(\nabla_{\mathbf{x}_u} \mathcal{L}(\mathbf{x}_u, \mathbf{q}^\star, \boldsymbol{\theta})\right), $ 
where $\epsilon$ controls the perturbation magnitude under some suitable power
constraint with respect to the original RSSI values. The adversarial input computed as $
\mathbf{x}_{\mathrm{adv}, u} = \mathbf{x}_u + \boldsymbol{\delta}_u$ is used to assess the classifier's robustness to outlier inputs.

To characterize the credibility estimates, we adopt the standard reliability diagrams \cite{RD}  to visualize the calibration of credibility scores. Reliability diagrams are histograms presenting accuracy as a function of credibility estimates of the model's prediction. The reliability diagrams bin the classifier's credibility score into $S$ intervals of equal size. A test data point  $(\hat{\mathbf{x}}_u, q)$ from the test dataset $ ({X^{\mathrm{te}} }, {Q^{\mathrm{te}}})$ is placed in bin $\mathcal{B}_s$ if the model’s credibility on $\hat{\mathbf{x}}_u$ is contained within the bin, i.e., $(\hat{\mathbf{x}}_u, q) \in \mathcal{B}_s$ . For each bin $\mathcal{B}_s$, the within-bin accuracy is defined as:
\begin{equation}
\operatorname{Acc}\left(\mathcal{B}_s\right)=\frac{1}{\left|\mathcal{B}_s\right|} \sum_{(\hat{\mathbf{x}}_u, q) \in \mathcal{B}_s} \mathbbm{1}_{\{\hspace{0.05cm} \arg \max_{j \in \left\{1,2,\dots,Q\right\}} p_j(\hat{\mathbf{x}}_u) = {q}\}},
\end{equation}
which measures the fraction of test samples within the bin that are correctly classified.


It is noteworthy that rather than blindly trusting the model's predictions, the proposed approach enhances explainability by using nearest neighbors to provide example-based insights, aligning the DNN's intermediate computations with its output for more transparent decision-making

\section{Simulation Results}\label{sec:simulation}
In this section, we provide details of the simulation setup, dataset acquisition, and the DL model architecture, followed by a discussion of the results.
\subsection{Simulation Setup}
To simulate the BS-UE mmWave communications, we adopt
an urban scenario comprising the downtown sector of Boston with both line-of-sight (LOS) and non-line-of-sight (NLOS) users. The BS employs a ULA with $N_{\mathrm{BS}}=32$ antennas and is placed at a height of 15 meters while oriented towards the negative $y$-axis. The BS communicates with single-antenna UEs with a height of 2 meters. The service area measures 200 meters by 230 meters and is discretized into a user grid with a spacing of 0.37 meters.  Based on these configurations, a total number of 98,650 downlink UE channels are generated.  We construct the channel matrix for every UE position using the DeepMIMO dataset generator \cite{deepmimo2024}  according to the specified parameters and system configuration summarized in Table \ref{tab:hyperparameters}. To enhance the stability and efficiency of training, the channel vectors, the channel vectors  $\mathbf{h} \in \mathbb{C}^{N_{\mathrm{BS}} \times 1}$  are normalized by the largest absolute value in the channel's matrix.
All simulations are performed on a $10$-Core Intel(R) Xenon(R) Silver $4114$, $2.2 \mathrm{GHz}$ system equipped with an Nvidia Quadro P2000 graphics processing unit (GPU).

The BS performs analog beamforming using $M_{w}$ sensing beam using a $N_{\mathrm{BS}}$-DFT codebook, whereas, for the narrow beam codebook $\mathbf{W}$, we use an O-DFT  with OS factor of 4 to get a total of 128 narrow beams. The parameters of the neural network $\boldsymbol{\theta}$  are learned from a labeled dataset 
$\mathcal{D}=\left\{\left(\mathbf{x}_{u,k}, \mathrm{q}_k^{\star}\right): k=1, . ., K\right\}$ which is composed of $K$ labeled training samples. Each sample consists of the received power values as the input features and the O-DFT beam index as the target label generated using (\ref{beam_selection}). In all experiments, 70\% of the data is allocated for training, 10\% for validation, and the remaining 20\% for testing. The calibration dataset is created by reserving a portion of the test data not utilized for evaluation.

\begin{table}[!t]
\centering
\caption{\textsc{Hyper-parameters for channel generation}}
\label{tab:hyperparameters}
\begin{adjustbox}{width=0.6\columnwidth,center}\footnotesize
\begin{tabular}{|c|c|}
\hline
\textbf{Name of scenario} & \textbf{Boston-5G} \\ \hline
Active BS          & 1               \\ \hline
BS transmit power          & 30 dBm               \\ \hline
Active users       & 900-1622          \\ \hline
Number of antennas (x, y, z) & (1, 32, 1)  \\ \hline
Carrier frequency    & 28 GHz        \\ \hline
System bandwidth          & 0.5 GHz        \\ \hline
Antenna spacing    & 0.5            \\ \hline
OFDM sub-carriers  & 512              \\ \hline
OFDM sampling factor & 1              \\ \hline
OFDM limit         & 1              \\ \hline
Number of multi-paths & 5              \\ \hline
\end{tabular}
\end{adjustbox}
\end{table}

\begin{table}[!t] 
\centering
\caption{\textsc{Architecture and Training Hyper-parameters}}
\label{tab:CNN-parameters}
\begin{adjustbox}{width=0.7\columnwidth,center}\footnotesize
\begin{tabular}{|c|c|c|c|c|} 
\hline
\textbf{Layer}   & \textbf{Filters} & \textbf{Kernel} & \textbf{Stride} & \textbf{Padding} \\ \hline
Conv 1           & 32               & 3$\times$3      & 1               & 1                 \\ \hline
Conv 2           & 64               & 3$\times$3      & 1               & 1                 \\ \hline
Conv 3           & 128              & 1$\times$1      & 1               & 0                 \\ \hline
\multicolumn{1}{|c|}{Fully connected} & \multicolumn{4}{c|}{ 128 units} \\ \hline
\multicolumn{2}{|c|}{Activation} & \multicolumn{3}{c|}{ ReLU (after each Conv layer)} \\ \hline
\multicolumn{2}{|c|}{Optimizer } & \multicolumn{3}{c|}{Adam, $10^{-3}$} \\ \hline
\multicolumn{2}{|c|}{Loss function } & \multicolumn{3}{c|}{Cross entropy} \\ \hline
\multicolumn{2}{|c|}{Epochs } & \multicolumn{3}{c|}{100} \\ \hline
\end{tabular}
\end{adjustbox}

\end{table}

\subsection{Deep Learning Model Architecture}
The proposed  DkNN beam classifier is implemented using a CNN, owing to its 
powerful feature extraction capabilities \cite{CNN_reason}. The designed classifier has $L=$ 4 layers with three convolution layers stacked with a fully connected layer. Each convolution layer is followed by the rectified linear unit (ReLU) activation to provide non-linearity to the convolutional layers. The fully connected layer takes the flattened input vector of local information and outputs the softmax probability distribution. For a quick nearest neighbor search on DkNN, we use an LSH from the FALCONN Python library \cite{andoni2015practical} and employ a grid parameter search to configure the number of neighbors to $\Tilde{k}=$ 10.  The hyperparameters of the DkNN beam classifier are summarized in Table \ref{tab:CNN-parameters}.

 \subsection{ Performance Evaluation}
For comparison purposes, we consider the following benchmark methods: 1) the upper bound beamforming based on the SVD of perfectly known channels under
$4$-bit quantized phase shifter \cite{heath2016overview}; 2) the DFT-based codebook
scanning directions with $N_{\mathrm{BS}}$ candidate beams at the BS
\cite{CI0}; and 3) the
O-DFT  codebook \cite{OS_Asmaa} with an oversampling factor ($OS$) = 4  and $ N_{\mathrm{BS}} \times$4 candidate beams at the BS. It is worth mentioning that the SVD upper bound
can only be reached when each user’s perfect channel is known at the BS. To evaluate the explainability and robustness of the proposed DkNN-based beam classifier, we compare its credibility estimates with the outputs of a standard softmax classifier, typically interpreted as
 model’s confidence estimates\cite{Deepknn}, with both classifiers using the same DNN architecture.

 To assess the optimal beam alignment accuracy, we use the \textit{top-k} accuracy metric, defined as the proportion of test samples where the optimal beam index falls within the top $k$ predicted beams.

\subsubsection{Measurement Noise Analysis}
\begin{figure}[t]
        \centering
         \includegraphics[width=0.75\linewidth]{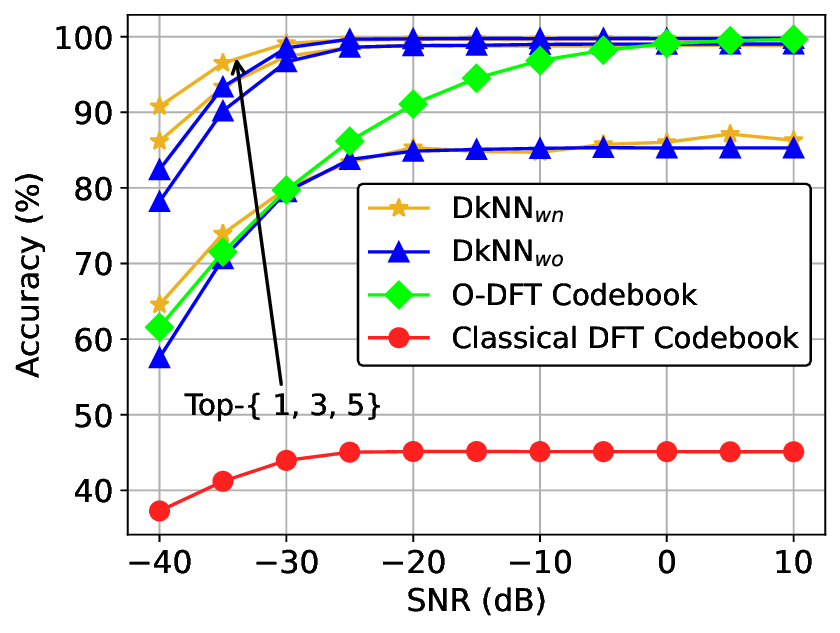}
        
    \caption{Accuracy versus SNR values (dB) for different schemes.}
    \label{fig:combined_acc}
    \vspace{-0.5cm}
\end{figure}

The accuracy of beam prediction is influenced by the measured power of the beamforming signals, where noise in these signals can significantly affect beam alignment performance. We consider two different cases: 1) \textit{DkNN trained without measurement noise} ($\text{DkNN}_{\text{wo}}$) where the CNN model  is trained with no noise
present in its training data, but it is then exposed to
noisy signal during the testing/validation stage; 2) \textit{DkNN trained with measurement noise} ($\text{DkNN}_{\text{wn}}$)  where the CNN model is trained with the expected measurement noise and is then deployed in a network with the
expected noise distribution.

Fig. \ref{fig:combined_acc} compares the accuracy performance of different schemes across different SNR values. In our analysis, the noise power ranges considered while generating the training
data are between -28dBm to -90dBm.   It is evident that $\text{DkNN}_{\text{wn}}$ maintains higher accuracy across expected noise levels compared to $\text{DkNN}_{\text{wo}}$. In the NLOS environment, the DKNN trained for expected noise maintains top-5 and top-3 accuracy above 92\% until SNR drops below -35 dB, while $\text{DkNN}_{\text{wo}}$ sees a more significant decline for higher noise levels. The reason for the improved performance of the
DkNN is that it can train the neural network with the received signal
containing channel measurement noise and is more robust at low SNR values. In comparison, the O-DFT (x4) codebook-based solution is less robust against noise, with a steady decrease in accuracy below -10dB SNR. On the other hand, the classical DFT codebook solution shows minimal adaptability, remaining below 45\% accuracy across all SNR levels, illustrating its limitations in dynamic environments and susceptibility to measurement noise.

\begin{figure}[t]
    \centering
        \includegraphics[width=0.75\linewidth]{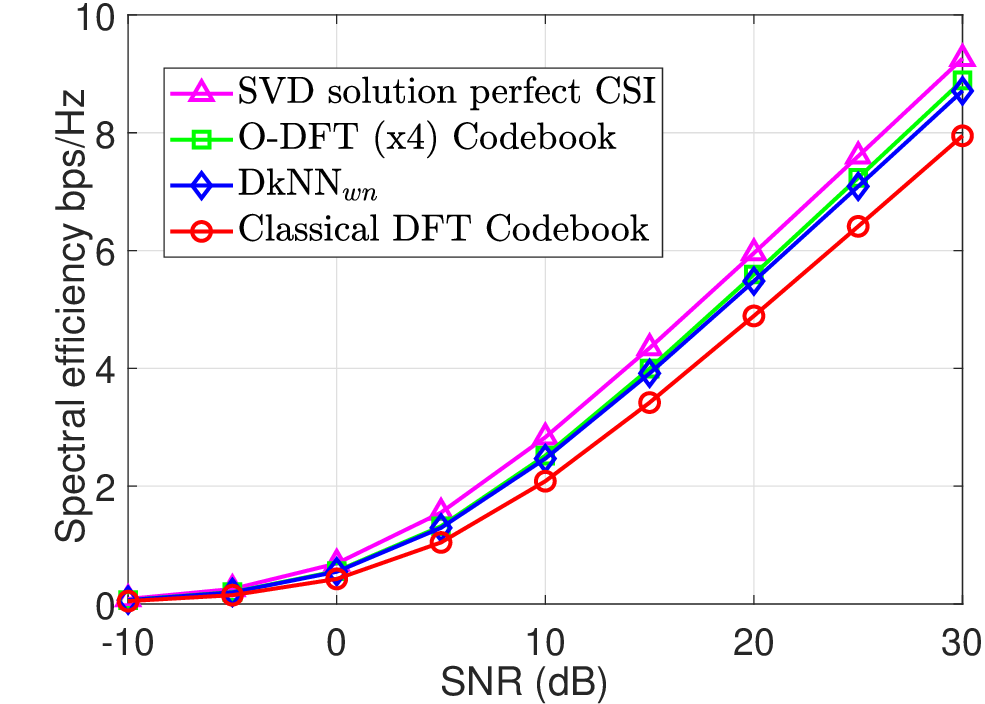}
    \caption{Spectral efficiency versus the SNR for different schemes.}
    \label{fig:combined_SE}
        \vspace{-0.5cm}
\end{figure}

Fig. \ref{fig:combined_SE} illustrates the spectral efficiency versus SNR for various beam alignment methods. The proposed $\text{DkNN}_{\text{wn}}$-based algorithm exhibits resilience against measurement noise and achieves approximately 98.5\% of the 128-DFT codebook performance in terms of spectral efficiency at all SNR values while reducing the beam training overhead by $\sim$ 75\% compared to the O-DFT codebook solution and significantly outperforming classical DFT codebook based solution.  Compared to the O-DFT codebook-based solutions, which require an exhaustive
beam search over 128 beams, the proposed method requires beam sweeping with only a subset of $N_{\mathrm{BS}}$ beams and an additional \textit{top-k} predicted beams.

\subsubsection{Explainability and Robustness Evaluation}
We evaluate the explainability and robustness of the proposed DkNN-based beam classifier by comparing its prediction credibility to that of the softmax classifier. The softmax classifier estimates confidence using output probability distributions but lacks explainability and robustness. In contrast, the DkNN provides interpretability through nearest neighbors, offering human-understandable explanations for intermediate computations at each layer, making it a valuable debugging tool. We show that the softmax classifier is poorly calibrated and overestimates confidence when predicting out-of-distribution inputs.

\begin{figure}[t]
    \centering
    \begin{subfigure}[t]{1\linewidth}
        \centering
        \includegraphics[width=0.75\linewidth]{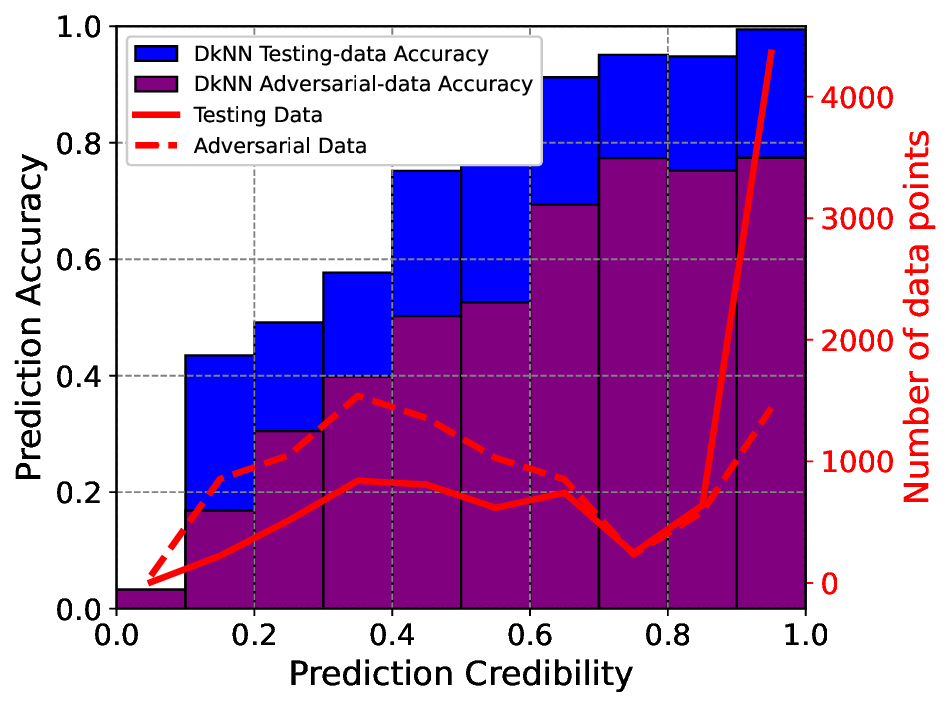}
        \caption{Reliability diagram for DkNN Classifier}
        \label{fig:RD1}
    \end{subfigure}
    \hfill
    \begin{subfigure}[t]{1\linewidth}
        \centering
          \includegraphics[width=0.75\linewidth]{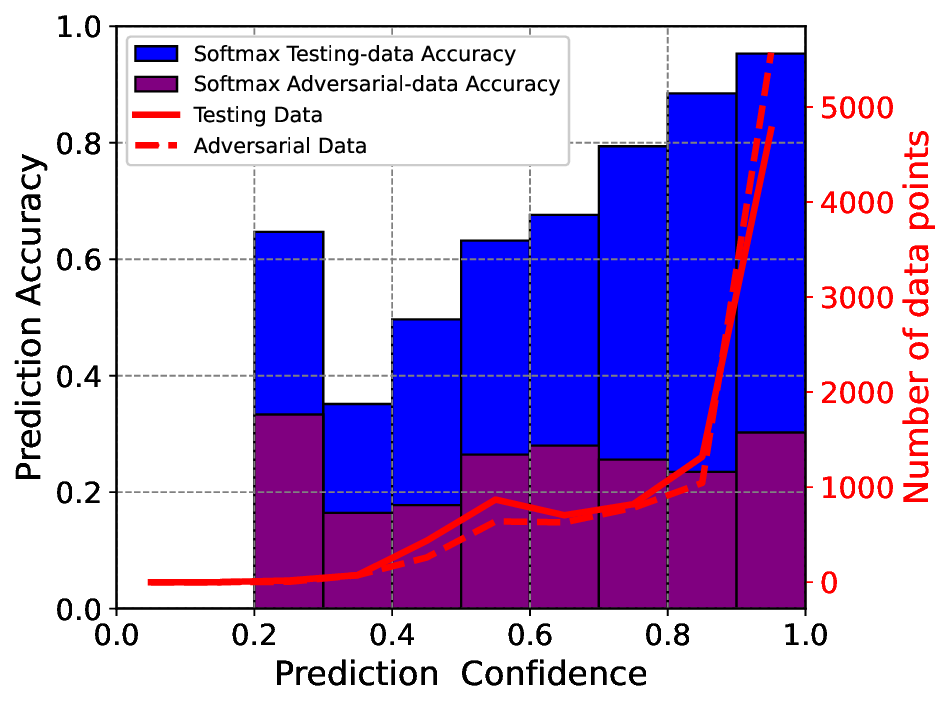}
        \caption{Reliability diagram for Softmax Classifier}
        \label{fig:RD2}
    \end{subfigure}
    \caption{Reliability diagrams for the classifiers on the Boston-5G dataset.}
    \label{fig:RD}
  
\end{figure}

\vspace{-0.01cm}
Figure \ref{fig:RD} presents the reliability diagrams for the DkNN and the naive softmax classifier. The distribution of credibility/confidence values across the data is given by the number of data points in each credibility bin, reflected by the red line overlaid on the bars. The softmax classifier lacks calibration as it consistently exhibits high confidence on both test and adversarial data, making it ineffective in identifying outliers.  Figure \ref{fig:RD1} demonstrates that the proposed DkNN classifier exhibits superior calibration by assigning low credibility to adversarial samples, effectively filtering outliers. It achieves 5$\times$ and 3.5$\times$ robustness improvements at credibility thresholds of 0.2 and 0.4, respectively. Figure \ref{fig:RD2} illustrates the softmax classifier reliability diagram, which shows overconfidence, misclassifying more than 80\% of adversarial examples with high confidence ($>$0.9). Note that because of the NLOS environment in the Boston-5G dataset,  the test dataset contains many test inputs that are
difficult to classify, reflected by the lower mean accuracy of the
underlying CNN. Still, the DkNN recovers some accuracy on adversarial examples by leveraging representations from CNN's internal layers and, therefore, is better calibrated than its softmax equivalent: its reliability diagrams follow the linear relation between accuracy and CNN's credibility/confidence.

\section{Conclusion}\label{sec:conclusion}
In this paper, we have developed an explainable and robust beam alignment framework for  6G mmWave networks. The proposed solution trains a custom-designed CNN-based BAE by collecting RSSI measurements over a small number of compact wide sensing beams from the DFT codebook to select the best narrow beam from the O-DFT codebook for IA and data transmission. To improve the explainability and robustness of the trained BAE, we have proposed a model-agnostic  DL-aided framework utilizing the  DkNN algorithm, which inspects the internal representations of the model to evaluate how well their predictions conform with the training data, providing
interpretable insights into beam selection and robustness against outlier inputs. Compared to the classical DFT codebook-based solutions, the proposed approach reduces beam training overhead by 75\% while achieving near-optimal accuracy. Moreover, the proposed DkNN-based algorithm effectively filters outlier inputs and  provides interpretable insights rationalizing the beam prediction decisions, enhancing both explainability and robustness. Future works could involve benchmarking the proposed solution against other explainable AI methods and assessing the impact of different adversarial attacks on the robustness of beam prediction.

\bibliographystyle{IEEEtran}\bibliography{XAI-bibliography}

\end{document}